\documentclass[twocolumn,showpacs,preprintnumbers,amsmath,amssymb, showkeys]{revtex4}

\usepackage{graphicx}
\usepackage{dcolumn}
\usepackage{bm}
\usepackage{epsf}

\begin{document}


\title{External field induced switching of tunneling current in
the coupled quantum dots}

 \author{V.\,N.\,Mantsevich}
 \altaffiliation{vmantsev@spmlab.phys.msu.ru}
\author{N.\,S.\,Maslova}%
 \email{spm@spmlab.phys.msu.ru}
 \author{P.\,I.\,Arseyev}
 \altaffiliation{ars@lpi.ru}
\affiliation{Moscow State University, Department of  Physics, 119991
Moscow, Russia\\~\\ P.N. Lebedev Physical institute of RAS, 119991,
Moscow, Russia}

\date{\today }
6 pages, 2 figures
\begin{abstract}
We investigated the tunneling current peculiarities in the system of
two coupled by means of the external field quantum dots (QDs) weakly
connected to the electrodes in the presence of Coulomb correlations.
It was found that tuning of the external field frequency induces
fast multiple tunneling current switching and leads to the negative
tunneling conductivity. Special role of multi-electrons states was
demonstrated. Moreover we revealed conditions for bistable behavior
of the tunneling current in the coupled QDs with Coulomb
correlations.
\end{abstract}

\maketitle

\section{Introduction}

Electron tunneling through the system of coupled quantum dots in the
presence of strong Coulomb correlations seems to be one of the most
interesting and important problems in the physics of nanostructures.
Tunneling current changes localized states electron filling numbers,
consequently, the spectrum and electron density of states are also
modified due to Coulomb interaction of localized electrons.

The present day experimental technique gives possibility to create
QDs with a given set of parameters and to produce coupled QDs with
different spatial geometries \cite{Vamivakas},\cite{Munoz-Matutano},
which give an opportunity to analyze non-equilibrium and
non-stationary effects in the small size correlated structures
\cite{Goldin},\cite{Kikoin},\cite{Mantsevich},\cite{Mantsevich_1},\cite{Mantsevich_2},\cite{Paaske},\cite{Kaminski}.
Thereby the main effort in the physics of QDs is devoted to the
investigation of non-equilibrium charge states and different spin
configurations due to the electrons tunneling
\cite{Kikoin_1},\cite{Orellana},\cite{Mantsevich_3} through the
system of coupled QDs in the presence of strong Coulomb interaction.

Double QDs systems behavior is recently under careful investigation
because of the variable inter-dot tunneling coupling
\cite{Oosterkamp},\cite{Livermore}, which is the physical reason for
non-linearity formation and consequently for existence of such
phenomena as bifurcations \cite{Rotter},\cite{Mantsevich_4} and
bistability \cite{Orellana},\cite{Goldman}. That's why double QDs
can be applied for logic gates fabrication based on the effect of
ultra-fast switching between intrinsic stable states.

In the present paper we consider electron tunneling through the QDs
with Coulomb correlations in the regime when coupling between the
dots is carried out by means of the external field with Rabi
frequency $\Omega$. We analyzed tunneling current behavior in terms
of pseudo operators with constraint
\cite{Coleman},\cite{Coleman_1},\cite{Wingreen},\cite{Mantsevich_3}.
For large values of applied bias Kondo effect is not essential so we
neglect any correlations between electron states in the QDs and in
the leads. This approximation allows to describe correctly
non-equilibrium occupation of any single- and multi-electron state
due to  the tunneling processes.

We revealed the presence of negative tunneling conductivity in
certain ranges of the applied bias voltage and analyzed the multiple
tunneling current switching caused by the external field frequency
tuning.

\section{Model}
We consider a system of coupled QDs with the single particle levels
$\widetilde{\varepsilon}_{1}$ è $\widetilde{\varepsilon}_{2}$
connected to the two leads. The Hamiltonian can be written as:

\begin{eqnarray}
\hat{H}&=&\sum_{\sigma}c_{1\sigma}^{+}c_{1\sigma}\widetilde{\varepsilon}_{1}+
\sum_{\sigma}c_{2\sigma}^{+}c_{2\sigma}\widetilde{\varepsilon}_{2}+
U_1\widehat{n}_{1\sigma}\widehat{n}_{1-\sigma}+\nonumber\\&+&U_2\widehat{n}_{2\sigma}\widehat{n}_{2-\sigma}+
\sum_{\sigma}\frac{\Omega}{2}(c_{1\sigma}^{+}c_{2\sigma}+c_{2\sigma}^{+}c_{1\sigma})
\end{eqnarray}

where operator $c_{l\sigma}$ creates an electron in the dot $i$ with
spin $\sigma$, $\widetilde{\varepsilon}_{l}$ is the energy of the
single electron level in the dot $i$ and inter-dot coupling is
realized by means of the external field with Rabi frequency
$\Omega$, $n_{l\sigma}=c_{l\sigma}^{+}c_{l\sigma}$ and $U_{1(2)}$ is
the on-site Coulomb repulsion of localized electrons. When the
coupling between QDs exceeds the value of interaction with the
leads, one has to use the basis of exact eigenfunctions and
eigenvalues of the coupled QDs without interaction with the leads.
In this case all energies of single- and multi-electron states are
well known:

One electron in the system: two single electron states with the wave
function

\begin{eqnarray}
\psi_{i}^{\sigma}=\mu_{i}\cdot|0\uparrow\rangle|00\rangle+\nu_{i}\cdot|00\rangle|0\uparrow\rangle
\end{eqnarray}
Single electron energies and coefficients $\mu_{i}$ and $\nu_{i}$
can be found as an eigenvalues and eigenvectors of matrix:

\begin{eqnarray}
\begin{pmatrix}
\varepsilon_{1} && -\frac{\Omega}{2}\\
-\frac{\Omega}{2} && \varepsilon_{2}
\end{pmatrix}
\label{m1}
\end{eqnarray}

Two electrons in the system: two states with the same spin
$\sigma\sigma$ and $-\sigma-\sigma$ and four two-electron states
with the opposite spins $\sigma-\sigma$ with the wave function:

\begin{eqnarray}
\psi_{j}^{\sigma-\sigma}&=&\alpha_{j}\cdot|\uparrow\downarrow\rangle|00\rangle+\beta_{k}\cdot|\downarrow0\rangle|0\uparrow\rangle+\nonumber\\&+&
\gamma_{j}\cdot|0\uparrow\rangle|\downarrow0\rangle+\delta_{j}\cdot|00\rangle|\uparrow\downarrow\rangle\nonumber\\
\end{eqnarray}

Two electron energies and coefficients $\alpha_{j}$, $\beta_{j}$,
$\gamma_{j}$ and $\delta_{j}$ are the eigenvalues and eigenvectors
of matrix:

\begin{eqnarray}
\begin{pmatrix}2\varepsilon_{1}+U_{1} && -\frac{\Omega}{2} && -\frac{\Omega}{2} && 0 \\
-\frac{\Omega}{2} && \varepsilon_{1}+\varepsilon_{2} && 0 && -\frac{\Omega}{2}\\
-\frac{\Omega}{2} && 0 && \varepsilon_{1}+\varepsilon_{2} && 0\\
0 && -\frac{\Omega}{2} && -\frac{\Omega}{2} &&
2\varepsilon_{2}+U_{2}\end{pmatrix} \label{m2}\end{eqnarray}

Three electrons in the system: two three-electron states with the
wave function

\begin{eqnarray}
\psi_{m}^{\sigma\sigma-\sigma}&=&p_{m}|\uparrow\downarrow\rangle|\uparrow\rangle+q_{m}|\uparrow\rangle|\uparrow\downarrow\rangle\nonumber\\
m&=&\pm1
\end{eqnarray}

Three electron energies and coefficients $p_{m}$ and $Q_{m}$ can be
found as an eigenvalues and eigenvectors of matrix:

\begin{eqnarray}
\begin{pmatrix}2\varepsilon_{1}+\varepsilon_{2}+U_{1} && -\frac{\Omega}{2}\\
-\frac{\Omega}{2} &&
2\varepsilon_{2}+\varepsilon_{1}+U_{2}\end{pmatrix}
\label{m3}\end{eqnarray}

Four electrons in the system: one four-electron state with energy
$E_{IVl}=2\varepsilon_1+2\varepsilon_2+U_1+U_{2}$ and wave function

\begin{eqnarray}
\psi_{l}=|\uparrow\downarrow\rangle|\uparrow\downarrow\rangle
\end{eqnarray}

If coupled QDs are connected with the leads of the tunneling contact
the number of electrons in the dots changes due to the tunneling
processes. Transitions between the states with different number of
electrons in the two interacting QDs can be analyzed in terms of
pseudo-particle operators with constraint on the physical states
(the number of pseudo-particles). Consequently, the electron
operator $c_{\sigma l}^{+}$  $(l=1,2)$ can be written in terms of
pseudo-particle operators as:

\begin{eqnarray}
c_{\sigma l}^{+}&=&\sum_{i}X_{i}^{\sigma l}f_{\sigma
i}^{+}b+\sum_{j,i}Y_{ji}^{\sigma-\sigma
l}d_{j}^{+\sigma-\sigma}f_{i-\sigma}+\\&+&\sum_{j,i}Y_{i}^{\sigma\sigma
l}d^{+\sigma\sigma}f_{i\sigma}+\sum_{m,j}Z_{mj}^{\sigma\sigma-\sigma
l}\psi_{m-\sigma}^{+}d_{j}^{\sigma-\sigma}+\nonumber\\&+&\sum_{m}Z_{m}^{\sigma-\sigma-\sigma
l}\psi_{m\sigma}^{+}d^{-\sigma-\sigma}+\sum_{m}W_{m}^{\sigma-\sigma-\sigma
l}\varphi^{+}\psi_{m\sigma}\nonumber\
\end{eqnarray}

where $f_{\sigma}^{+}(f_{\sigma})$ and
$\psi_{\sigma}^{+}(\psi_{\sigma})$- are pseudo-fermion creation
(annihilation) operators for the electronic states with one and
three electrons correspondingly. $b^{+}(b)$,
$d_{\sigma}^{+}(d_{\sigma})$ and $\varphi^{+}(\varphi)$- are slave
boson operators, which correspond to the states without any
electrons, with two electrons or four electrons. Operators
$\psi_{m-\sigma}^{+}$- describe system configuration with two spin
up electrons $\sigma$ and one spin down electron $-\sigma$ in the
symmetric and asymmetric states.

Matrix elements $X_{i}^{\sigma l}$, $Y_{ji}^{\sigma-\sigma l}$,
$Y_{ji}^{\sigma\sigma l}$, $Z_{mj}^{\sigma\sigma-\sigma l}$,
$Z_{mj}^{\sigma-\sigma-\sigma l}$ and $W_{m}^{\sigma-\sigma-\sigma
l}$ can be evaluated as:

\begin{eqnarray}
X_{i}^{\sigma l}&=&\langle\psi_{i}^{\sigma}|c_{\sigma l}^{+}|0\rangle\nonumber\\
Y_{ji}^{\sigma-\sigma l}&=&\langle\psi_{j}^{\sigma-\sigma}|c_{\sigma il}^{+}|\psi_{i}^{-\sigma}\rangle\nonumber\\
Y_{ji}^{\sigma\sigma l}&=&\langle\psi_{j}^{\sigma\sigma}|c_{\sigma l}^{+}|\psi_{i}^{\sigma}\rangle\nonumber\\
Z_{mj}^{\sigma\sigma-\sigma l}&=&\langle\psi_{m}^{\sigma\sigma-\sigma}|c_{\sigma l}^{+}|\psi_{j}^{\sigma-\sigma}\rangle\nonumber\\
Z_{m}^{\sigma-\sigma-\sigma l}&=&\langle\psi_{m}^{\sigma-\sigma-\sigma}|c_{\sigma l}^{+}|\psi^{-\sigma-\sigma}\rangle\nonumber\\
W_{m}^{\sigma-\sigma-\sigma
l}&=&\langle\psi_{l}^{\sigma\sigma-\sigma-\sigma}|c_{\sigma
l}^{+}|\psi_{m}^{\sigma-\sigma-\sigma}\rangle\
\end{eqnarray}

Finally one can easily express matrix elements through the matrixes
(\ref{m1}), (\ref{m2}), (\ref{m3}) eigenvectors elements:

\begin{eqnarray}
X_{i}^{\sigma 1}=\mu_{i};
X_{i}^{\sigma 2}=\nu_{i}\nonumber\\
Y_{ji}^{\sigma-\sigma 1}=\alpha_j\mu_i+\beta_j\nu_i\nonumber\\
Y_{ji}^{\sigma-\sigma 2}=\delta_j\nu_i+\gamma_j\mu_i\nonumber\\
Y_{ji}^{\sigma\sigma 1}=\nu_i;
Y_{ji}^{\sigma\sigma 2}=\mu_i\nonumber\\
Z_{mj}^{\sigma\sigma-\sigma 1}=p_m\gamma_j+q_m\delta_j\nonumber\\
Z_{mj}^{\sigma\sigma-\sigma 2}=p_m\alpha_j+q_m\beta_j\nonumber\\
Z_{mj}^{\sigma-\sigma-\sigma 1}=p_m;
Z_{mj}^{\sigma-\sigma-\sigma 1}=q_m\nonumber\\
W_{m}^{\sigma-\sigma-\sigma 1}=q_m; W_{m}^{\sigma-\sigma-\sigma
2}=p_m
\end{eqnarray}

The constraint on the space of the possible system states have to be
taken into account:

\begin{eqnarray}
\widehat{n}_{b}+\sum_{i\sigma}\widehat{n}_{fi\sigma}+\sum_{j\sigma\sigma^{'}}\widehat{n}_{dj}^{\sigma\sigma^{'}}+\sum_{m\sigma}\widehat{n}_{\psi
m\sigma}+\widehat{n}_{\varphi}=1 \label{limit}
\end{eqnarray}

Condition (\ref{limit}) means that the appearance  of any two
pseudo-particles in the system simultaneously is impossible.

Electron filling numbers in the coupled QDs can be expressed in
terms of the pseudo-particles filling numbers:

\begin{eqnarray}
\widehat{n}_{\sigma}^{el}&=&\sum_{l}c_{\sigma l}^{+}c_{\sigma
l}=\sum_{i,l}|X_{i}^{\sigma
l}|^{2}\widehat{n}_{fi\sigma}+\sum_{i,j,l}|Y_{ji}^{\sigma-\sigma
l}|^{2}\widehat{n}_{dj}^{\sigma-\sigma}+\nonumber\\&+&\sum_{i,l}|Y_{ji}^{\sigma\sigma
l}|^{2}\widehat{n}_{dj}^{\sigma\sigma}+\sum_{m,j,l}|Z_{mj}^{\sigma\sigma-\sigma
l}|^{2}\widehat{n}_{\psi
m-\sigma}+\nonumber\\&+&\sum_{m,l}|Z_{mj}^{-\sigma-\sigma\sigma
l}|^{2}\widehat{n}_{\psi
m\sigma}+\sum_{m,l}|W_{m}^{\sigma-\sigma-\sigma
l}|^{2}\widehat{n}_{\varphi}
\end{eqnarray}

Consequently, the Hamiltonian of the system can be written in terms
of the pseudo-particle operators:

\begin{eqnarray}
\hat{H}&=&\hat{H_{0}}+\hat{H}_{tun}\\
\hat{H_{0}}&=&\sum_{i\sigma}\varepsilon_{i}f_{i\sigma}^{+}f_{i\sigma}+\sum_{j\sigma\sigma^{'}}E_{IIj}^{\sigma\sigma^{'}}d_{j}^{+\sigma\sigma^{'}}d_{j}^{\sigma\sigma^{'}}+\nonumber\\&+&\sum_{m\sigma}E_{III}^{m\sigma}\psi_{m\sigma}^{+}\psi_{m\sigma}+E_{IVl}\varphi_{\sigma}^{+}\varphi_{\sigma}+\nonumber\\&+&\sum_{k\sigma}(\varepsilon_{k\sigma}-eV)c_{k\sigma}^{+}c_{k\sigma}+\sum_{p\sigma}\varepsilon_{p\sigma}c_{p\sigma}^{+}c_{p\sigma}\nonumber\\
\hat{H}_{tun}&=&
\sum_{\sigma}\frac{\Omega}{2}(c_{1\sigma}^{+}c_{2\sigma}+c_{2\sigma}^{+}c_{1\sigma})+\nonumber\\&+&\sum_{k\sigma}T_{k}(c_{k\sigma}^{+}c_{\sigma1}+c_{\sigma1}^{+}c_{k\sigma})+(k\leftrightarrow
p;1\leftrightarrow2)\nonumber\
\end{eqnarray}

where $\varepsilon_i$, $E_{IIj}^{\sigma\sigma^{'}}$,
$E_{III}^{m\sigma}$ and $E_{IVl}$-are the energies of the single-,
double-, triple- and quadri-electron states.
$\varepsilon_{k(p)\sigma}$-is the energy of the conduction electrons
in the states $k$ and $p$ correspondingly.
$c_{k(p)\sigma}^{+}/c_{k(p)\sigma}$ are the creation (annihilation)
operators in the leads of the tunneling contact. $T_{k(p)}$-are the
tunneling amplitudes, which we assume to be independent on momentum
and spin. Indexes $k(p)$ mean only that tunneling takes place from
the system of coupled QDs to the conduction electrons in the states
$k$ and $p$ correspondingly.

Bilinear combinations of pseudo-particle operators are closely
connected with the density matrix elements. So, similar expressions
can be obtained from equations for the density matrix evolution but
method based on the pseudo particle operators is more compact and
convenient. The tunneling current through the proposed system
written in terms of the pseudo-particle operators has the form:

\begin{eqnarray}
\widehat{I}_{k\sigma}&=&\sum_{k}\frac{\partial
\widehat{n}_{k}}{\partial t}=i [\sum_{ik}X_{i}^{\sigma 1}T_{k}
c_{k\sigma}f_{i\sigma}^{+}b+\nonumber\\&+&\sum_{ijk}Y_{ji}^{\sigma-\sigma
1}T_{k}
c_{k\sigma}d_{j}^{+\sigma-\sigma}f_{i-\sigma}+\nonumber\\&+&\sum_{ijk}Y_{ji}^{\sigma\sigma
1}T_{k} c_{k\sigma}d_{j}^{+\sigma\sigma}f_{i\sigma}+\nonumber\\&+&
\sum_{mjk}Z_{mj}^{\sigma\sigma-\sigma
1}T_{k}c_{k\sigma}\psi_{m-\sigma}^{+}d_{j}^{\sigma-\sigma}+\nonumber\\&+&
\sum_{mjk}Z_{mj}^{-\sigma-\sigma\sigma
1}T_{k}c_{k\sigma}\psi_{m\sigma}^{+}d_{j}^{-\sigma-\sigma}+\nonumber\\&+&\sum_{mk}W_{m}^{\sigma-\sigma-\sigma
1}T_{k}c_{k\sigma}\varphi^{+}\psi_{m\sigma}-h.c.]
\end{eqnarray}

We set $\hbar=1$ and neglect changes in the electron spectrum and
local density of states in the tunneling contact leads, caused by
the tunneling current. Therefore equations of motion together with
the constraint on the space of the possible system states
(pseudo-particles number) (\ref{limit}) give the following
equations:

\begin{eqnarray}
Im\sum_{ik}T_{k}X_{i}^{\sigma 1}\cdot \langle
c_{k\sigma}f_{i\sigma}^{+}b\rangle=\nonumber\\
=\Gamma_{k}\sum_{i}[(1-n_{k\sigma}(\varepsilon_i))\cdot
n_{fi\sigma}-n_{k\sigma}(\varepsilon_{i})\cdot n_{b}](X_{i}^{\sigma 1})^{2}\nonumber\\
Im\sum_{ijk}Y_{ji}^{\sigma-\sigma 1}T_{k}\cdot \langle
c_{k\sigma}d_{j}^{+\sigma-\sigma}f_{i-\sigma}\rangle=\nonumber\\=
\Gamma_{k}\sum_{ij}[(1-n_{k\sigma}(E_{IIj}^{\sigma-\sigma}-\varepsilon_{i-\sigma}))\cdot
n_{dj}^{\sigma-\sigma}-\nonumber\\-n_{k\sigma}(E_{IIj}^{\sigma-\sigma}-\varepsilon_{i-\sigma})\cdot
n_{fi-\sigma}](Y_{ji}^{\sigma-\sigma
1})^{2}\nonumber\\
Im\sum_{ijk}Y_{ji}^{\sigma\sigma 1}T_{k}\cdot \langle
c_{k\sigma}d_{j}^{+\sigma\sigma}f_{i\sigma}\rangle=\nonumber\\=
\Gamma_{k}\sum_{ij}[(1-n_{k\sigma}(E_{IIj}^{\sigma\sigma}-\varepsilon_{i\sigma}))\cdot
n_{dj}^{\sigma\sigma}-\nonumber\\-n_{k\sigma}(E_{IIj}^{\sigma\sigma}-\varepsilon_{i\sigma})\cdot
n_{fi\sigma}](Y_{ji}^{\sigma\sigma
1})^{2}\nonumber\\
Im\sum_{mjk}Z_{mj}^{\sigma\sigma-\sigma 1} T_{k}\cdot \langle
c_{k\sigma}\psi_{m-\sigma}^{+}d_{j}^{\sigma-\sigma}\rangle=\nonumber\\=
\Gamma_{k}\sum_{mj}[(1-n_{k\sigma}(E_{III}^{m-\sigma}-E_{IIj}^{\sigma-\sigma}))\cdot
n_{\psi
m-\sigma}-\nonumber\\-n_{k\sigma}(E_{III}^{m-\sigma}-E_{IIj}^{\sigma-\sigma})\cdot
n_{dj}^{\sigma-\sigma}](Z_{mj}^{\sigma\sigma-\sigma
1})^{2}\nonumber\\
Im\sum_{mjk}Z_{mj}^{-\sigma-\sigma\sigma 1} T_{k}\cdot \langle
c_{k\sigma}\psi_{m\sigma}^{+}d_{j}^{-\sigma-\sigma}\rangle=\nonumber\\=
\Gamma_{k}\sum_{mj}[(1-n_{k\sigma}(E_{III}^{m\sigma}-E_{IIj}^{-\sigma-\sigma}))\cdot
n_{\psi
m\sigma}-\nonumber\\-n_{k\sigma}(E_{III}^{m\sigma}-E_{IIj}^{-\sigma-\sigma})\cdot
n_{dj}^{-\sigma-\sigma}](Z_{mj}^{-\sigma-\sigma\sigma
1})^{2}\nonumber\\
Im\sum_{mk}W_{m}^{\sigma-\sigma-\sigma 1} T_{k}\cdot \langle
c_{k\sigma}\varphi_{l}^{+}\psi_{m\sigma}\rangle=\nonumber\\
\Gamma_{k}\sum_{m}[(1-n_{k\sigma}(E_{IVl}-E_{III}^{m\sigma}))\cdot
n_{\varphi}-\nonumber\\-n_{k\sigma}(E_{IVl}-E_{III}^{m\sigma})\cdot
n_{\psi m\sigma}](W_{m}^{\sigma-\sigma-\sigma 1})^{2}
\label{tunneling_current}
\end{eqnarray}

Tunneling current $I_{k\sigma}$ is determined by the sum of the
right hand parts of the equations (\ref{tunneling_current}). Pseudo
particle filling numbers  $n_{fi}$, $n_{dj}^{\sigma-\sigma}$,
$n_{d}^{\sigma\sigma}$, $n_{\psi m}$ and $n_{\varphi}$ can be easily
obtained from the stationary linear system of equations
\cite{Mantsevich_3}.

\section{Results and discussion}

\begin{figure*} [h!]
\includegraphics[width=170mm]{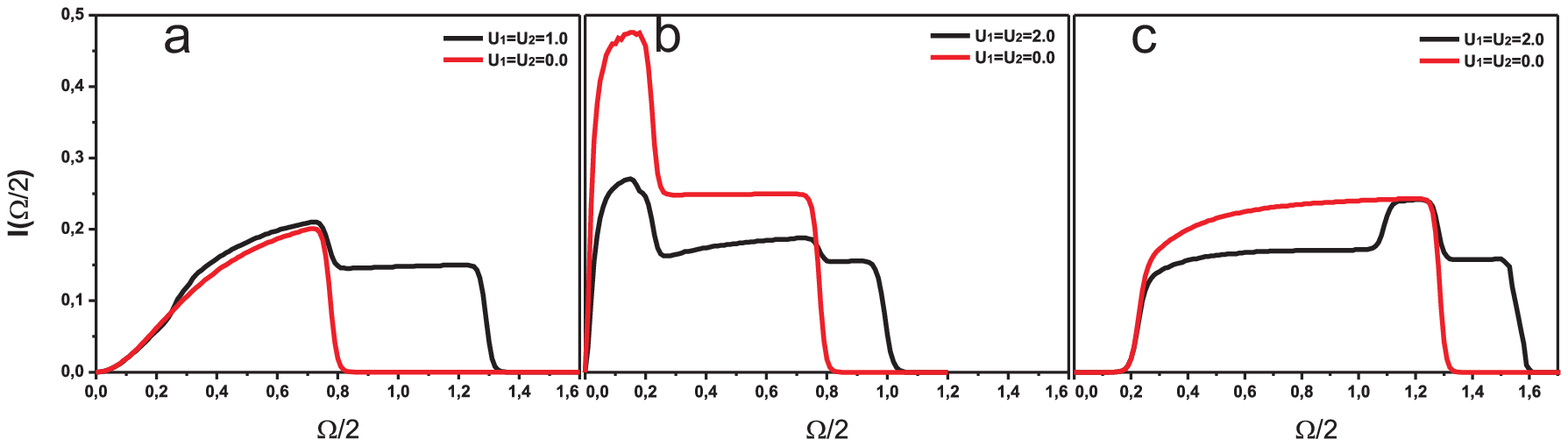}
\caption{Fig.1 (Color online) Tunneling current as a functions of
Rabi frequency in the case of symmetrical tunneling contact in the
presence (black line) and in the absence (red line) of Coulomb
interaction. a). $\varepsilon_{1}=1.20$, $\varepsilon_{2}=0.50$,
$eV=1.00$ $U_{1}=U_{2}=1.00$,$\Gamma_{k}=\Gamma_{p}=0.01$; b).
$\varepsilon_{1}=0.25$, $\varepsilon_{2}=0.20$, $eV=1.00$
$U_{1}=U_{2}=2.00$,$\Gamma_{k}=\Gamma_{p}=0.01$; c).
$\varepsilon_{1}=1.50$, $\varepsilon_{2}=1.10$, $eV=1.00$
$U_{1}=U_{2}=2.00$,$\Gamma_{k}=\Gamma_{p}=0.01$. } \label{Fig.1}
\end{figure*}

\begin{figure} [h!]
\includegraphics[width=70mm]{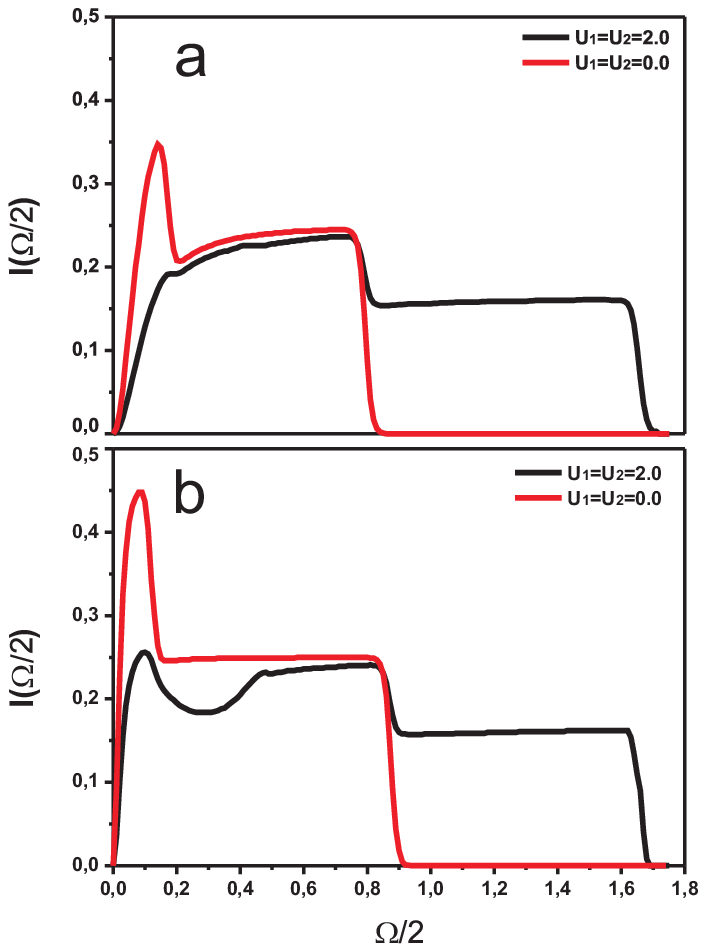}
\caption{Fig.2 (Color online) Tunneling current as a functions of
Rabi frequency in the case of symmetrical tunneling contact in the
presence (black line) and in the absence (red line) of Coulomb
interaction. a). $\varepsilon_{1}=0.90$, $\varepsilon_{2}=0.70$,
$eV=1.00$ $U_{1}=U_{2}=2.00$,$\Gamma_{k}=\Gamma_{p}=0.01$; b).
$\varepsilon_{1}=0.90$, $\varepsilon_{2}=0.85$, $eV=1.00$
$U_{1}=U_{2}=2.00$,$\Gamma_{k}=\Gamma_{p}=0.01$.} \label{Fig.2}
\end{figure}

The behavior of tunneling current with the  external field frequency
tuning for the different values of Coulomb interaction obtained from
equations (\ref{tunneling_current}) is depicted in
Fig.\ref{Fig.1}-Fig.\ref{Fig.2}. The general features of obtained
results is tunneling current switching due to the external field
frequency tuning and formation of negative tunneling conductivity.

We considered different experimental realizations: both
single-electron energy levels are situated between the  sample Fermi
level ($E_{F}=0$) and the value of applied bias voltage ($eV=1$ for
all the Figures) [Fig.\ref{Fig.1}b, Fig.\ref{Fig.2}]; both levels
exceed the sample Fermi level and the value of applied bias voltage
[Fig.\ref{Fig.1}c]; one of the energy levels is located between the
Fermi level and the value of applied bias voltage and another one
exceeds both of them [Fig.\ref{Fig.1}a].

External field frequency tuning results in the single-electron
energy levels spacing. When only one of the single-electron energy
levels is located between the Fermi level and the value of applied
bias voltage in the absence of Coulomb correlations [see red line in
Fig.\ref{Fig.1}a], tunneling current amplitude increases with the
increasing of external field frequency $\Omega$ until the lowest
energy level $\varepsilon_2$ continue being localized in the
$[E_{F};eV]$ energy gap. Further frequency growth leads to the
situation when both energy levels are localized out of the
$[E_{F};eV]$ energy interval and, consequently, sudden switching
$"$off$"$ of the tunneling current occurs. More complicated system
behavior corresponds to the case when Coulomb correlations are
considered [see black line in Fig.\ref{Fig.1}a]. The decreasing of
tunneling current amplitude with the external field frequency
increasing is through one more stable state when the tunneling
current value continues to differ from zero even when both single
electron energy levels are localized out of the $[E_{F};eV]$ energy
gap. This is the direct manifestation of Coulomb correlations,
because due to the presence of Coulomb correlations the
multi-electron energy states are located in the $[E_{F};eV]$ energy
interval. Consequently, tunneling current decreasing reveals more
complicated behavior. Moreover, obtained results demonstrate the
negative tunneling conductivity, when tunneling current decreases
with increasing of the external field frequency.

Tunneling current evolution as a function of the external field
frequency in the case when both single-electron energy levels are
initially localized in the $[E_{F};eV]$ energy gap is presented in
the Fig.\ref{Fig.1}b. In the absence of Coulomb interaction one can
see the growth of the tunneling current amplitude with the
increasing of external field frequency $\Omega$ until both
single-electron energy levels are localized in the $[E_{F};eV]$
energy interval [see red line in Fig.\ref{Fig.1}b]. Frequency
increasing leads to the situation when only one energy level is
located in the $[E_{F};eV]$ energy interval and it corresponds to
the sudden decreasing of the tunneling current, which results in the
formation of $"$step-down$"$ in the $I-\Omega$ characteristic.
Further increasing of the external field frequency leads to the
situation when both single electron energy levels are localized out
of the $[E_{F};eV]$ energy gap, and, consequently, one more
$"$step-down$"$ appears. When Coulomb correlations are taken into
account [see black line in Fig.\ref{Fig.1}b], decreasing of the
tunneling current amplitude with the external field frequency growth
takes place through one more stable state when the tunneling current
value continues to differ from zero even when both single electron
energy levels are localized out of the $[E_{F};eV]$ energy gap due
to the multi-electron energy states contribution. Consequently,
three $"$steps-down$"$ are visible in the $I-\Omega$ characteristic.

Tunneling current evolution in the case when both single-electron
energy levels are initially localized above the $[E_{F};eV]$ energy
gap is shown in the Fig.\ref{Fig.1}c. The lowest energy level
$\varepsilon_2$ is very close to $eV$. In this case tunneling
current value is equal to zero until both single-electron energy
levels are located above $eV$. In the absence of Coulomb interaction
one can clearly see one $"$step-up$"$ and one $"$step-down$"$ in the
$I-\Omega$ characteristic [see red line in Fig.\ref{Fig.1}c].
Switching $"$on$"$ of the tunneling current ($"$step-up$"$) takes
place when the external field frequency has the value, which is
enough for the lower energy level to be localized in the
$[E_{F};eV]$ energy interval. Switching $"$off$"$ ($"$step-down$"$)
takes place when frequency growth leads to the situation when both
energy levels are localized out of the $[E_{F};eV]$ energy interval.
$\varepsilon_1$ is higher than $eV$ and $\varepsilon_2$ is lower
than $E_{F}$. If one consider Coulomb correlations [see black line
in Fig.\ref{Fig.1}c], two $"$steps-up$"$ and two $"$steps-down$"$
are present in the $I-\Omega$ characteristic. This multiple
tunneling current switching $"$on$"$ and $"$off$"$ is the result of
multi-electron energy states contribution caused by the presence of
Coulomb correlations.

The other interesting effect associated with Coulomb correlations is
the presence of multi-stability in the coupled QDs for the
particular value of system parameters [see black line in
Fig.\ref{Fig.2}b]. Fig.\ref{Fig.2}b demonstrates that when both
single-electron energy levels are located slightly above $eV$ and
are close to each other, single value of the tunneling current
amplitude corresponds to the two values of external field frequency
$\Omega$ [see black line in Fig.\ref{Fig.2}b]. This effect
disappears when the single-electron energy levels spacing increases
[Fig.\ref{Fig.2}a] .

\section{Conclusion}

Tunneling through the system of two QDs with strong coupling between
localized electron states was analyzed by means of Heisenberg
equations for pseudo operators with constraint. Various
single-electron levels location relative to the sample Fermi level
and to the applied bias value in symmetric tunneling contact were
investigated.

We revealed the appearance of negative tunneling conductivity and
demonstrated multiple switching $"$on$"$ and $"$off$"$ of the
tunneling current depending on the Coulomb correlations value,
external field frequency amplitude and energy levels spacing. We
proved that Coulomb correlations strongly influence the system
behavior.

We demonstrated the presence of multi-stability in the coupled QDs
with Coulomb correlations when single value of the tunneling current
amplitude corresponds to the two values of external field frequency.

This work was partly supported by the RFBR and Leading Scientific
School grants. The support from the Ministry of Science and
Education is also acknowledged.


\pagebreak

\end{document}